\documentclass[12pt,a4paper,final]{iopart}

\usepackage{iopams}  

\expandafter\let\csname equation*\endcsname\relax
\expandafter\let\csname endequation*\endcsname\relax
\usepackage{amsmath}

\usepackage{graphicx}
\usepackage[breaklinks=true,colorlinks=true,linkcolor=blue,urlcolor=blue,citecolor=blue]{hyperref}

\newcommand{\pluscross}{\ensuremath{\small\{+,\times\small\}}}
\newcommand{\Msun}{\ensuremath{M_{ \odot }}}

\begin{document}

\title[Bumpy Black Holes]{Gravitational waves from extreme mass ratio inspirals around bumpy black holes}

\author{Christopher J. Moore}
\address{DAMTP, CMS, Wilberforce Road, Cambridge, CB3 0WA, UK}
\address{IST-CENTRA, Departamento de F{\'i}sica, Avenida Rovisco Pais 1, 1049 Lisboa, Portugal}
\ead{cjm96@cam.ac.uk}

\author{Alvin J. K. Chua}
\address{Institute of Astronomy, University of Cambridge, Madingley Road, Cambridge, CB3 0HA, UK}
\ead{ajkc3@ast.cam.ac.uk}

\author{Jonathan R. Gair}
\address{School of Mathematics, University of Edinburgh, The King's Buildings, Peter Guthrie Tait Road, Edinburgh, EH9 3FD, UK}
\ead{j.gair@ed.ac.uk}

\date{\today}

\begin{abstract}
The space based interferometer LISA will be capable of detecting the gravitational waves emitted by stellar mass black holes or neutron stars slowly inspiralling into the supermassive black holes found in the centre of most galaxies. The gravitational wave signal from such an extreme mass ratio inspiral (EMRI) event will provide a unique opportunity to test whether the spacetime metric around the central black hole is well described by the Kerr solution. In this paper a variant of the well studied ``analytic kludge'' model for EMRIs around Kerr black holes is extended to a family of parametrically deformed \emph{bumpy black holes} which preserve the basic symmetries of the Kerr metric. The new EMRI model is then used to quantify the constraints that LISA observations of EMRIs may be able to place on the deviations, or \emph{bumps}, on the Kerr metric.
\end{abstract}
 
\pacs{04.30.Db, 04.30.Tv, 04.50.Kd, 04.70.Bw, 04.80.Cc, 95.30.Sf}
\submitto{\CQG}
 
\section{Introduction}
The spectacular recent progress in the field of gravitational wave (GW) astronomy has opened up new possibilities for testing several key predictions of general relativity (GR). For example, the first detection of gravitational waves by Advanced LIGO \cite{PhysRevLett.116.061102,PhysRevLett.116.131103} was used to bound the Compton wavelength of the graviton and confirm that GW propagation was consistent with GR \cite{PhysRevLett.116.221101}. Evidence that GWs propagate at anything other than the speed of light would constitute a ``smoking gun'' for the failure of GR. As additional detectors, including Advanced Virgo \cite{2015CQGra..32b4001A} and KAGRA \cite{2012CQGra..29l4007S,2013PhRvD..88d3007A}, come online it will also become possible to distinguish different GW polarisation components in short-lived black hole (BH) binary merger signals\footnote{GW polarisation tests using just the two LIGO interferometers are possible for long-lived sources; for example, continuous GWs [7] or a stochastic GW background [8]. The daily rotation of the interferometers modulates the detector response differently for different GW polarisation states.}; if they reveal anything other than the transverse-traceless polarisations of GR, such as the ``breathing'' mode of scalar-tensor gravity (see e.g.\ \cite{1992CQGra...9.2093D}, and references therein), this would be similarly clear evidence for the failure of GR. 
 
This paper considers the possibility of testing another key prediction of GR together with the ``no-hair'' theorems: \emph{the spacetime around an astrophysical black hole is described by the Kerr metric} \cite{Kerr1963}. In contrast to the tests mentioned above, a deviation from the Kerr metric could either indicate a failure of GR, or it could point to a problem with the ``no-hair'' theorems and raise the possibility of exotic compact objects within GR. As a thought experiment, perhaps the most direct way to probe the metric around a BH would be to measure the geodesic trajectories of a large number of test particles; fortunately a practical test almost as direct will soon become possible.

The space-based GW detector LISA \cite{2017arXiv170200786A,PhysRevLett.116.231101}, scheduled for launch in the early 2030s, will observe the GWs from compact objects (COs) such as neutron stars or stellar mass BHs inspiralling into supermassive BHs in the mass range ${(10^{5}\textrm{--}10^{7})\Msun}$ \cite{Gairetal}. These events are known as extreme mass ratio inspirals (EMRIs). The majority of observed EMRI events are expected to be BH-BH mergers; this is partly due to mass segregation concentrating heavier BHs in the galactic centre and partly because their louder intrinsic amplitude enables them to be detected out to greater distances. The small CO approximates a test particle, and over the short orbital timescale follows a nearly geodesic trajectory in the background metric of the supermassive BH. The system radiates GWs at harmonics of the geodesic frequencies, so the GW frequency spectrum encodes details of the instantaneous geodesic trajectory. Over longer radiation reaction timescales the CO orbit changes (adiabatically, in the extreme mass ratio limit) as energy and angular momentum are radiated away. LISA will operate for several years, longer than the radiation reaction timescale, and will observe the CO orbit slowly traverse a one-parameter family of geodesics as the CO inspirals, and eventually plunges, into the central BH. It is believed that most EMRI events will be extremely ``clean'', meaning the systems consist solely of two BHs without the complicating effects of the object's internal structure, the presence of other perturbing bodies, or significant perturbation from an accretion disk (although see \cite{2008PhRvD..77j4027B,2011PhRvD..84b4032K,2007PhRvD..75f4026B}). It has long been realised that these features make EMRIs an ideal laboratory for mapping the metric around a supermassive BH \cite{1995PhRvD..52.5707R,Ryan} and testing strong field gravity (for a review, see \cite{2007CQGra..24R.113A}).
 
The number of EMRI that LISA will detect is highly uncertain; it is expected that between a few tens and a few thousands of events will exceed the threshold signal-to-noise ratio (SNR) for detection (this is itself uncertain, but lies in the range $15\!<\!\textrm{SNR}_{\textrm{Threshold}}\!<\!30$) with perhaps a few tens of ``golden EMRIs'' exceeding an SNR of 100 \cite{Gairetal,2017arXiv170504259C,2017arXiv170309722B}. This large SNR together with the fact that EMRI waveform exibit strong modulations due to relativistic precession (see section \ref{sec:deltadot}), which depends sensitively on the system parameters, means that it will be possible to measure those parameters very accurately (e.g.\ errors on the central BH mass and spin as low as a few parts in $10^{5}$ \cite{BarackCutler2008,2009PhRvD..79h4021H,2017arXiv170504259C}). It should also be possible to place comparably stringent constraints on any ``non-Kerr-ness'' of the metric.
 
One approach to testing the hypothesis that the metric is Kerr is to compare the observed EMRI waveforms against two sets of predictions: one calculated in the Kerr metric and a second in an alternative (e.g.\ the slowly rotating BH solution to dynamical Chern-Simons gravity \cite{CanizaresGairSopuerta} or a Kerr BH with a small anomalous quadrupole moment \cite{BarackCutler}). Such an approach is limited to metrics which have already been considered theoretically. A alternative, which avoids this limitation, is to construct a large family of metrics that are continuously parameterised deformations of the Kerr metric (or ``bumpy BHs'') and to place constraints on all of the various deformations (or ``bumps''). Some specific alternatives to the Kerr metric may be captured exactly by this family, but it is hoped that even those that are not will still give EMRI signals similar enough to a member of the family to reveal a deformation from Kerr. In this way a systematic and model independent test of the Kerr metric is possible. However, it should be noted that such a scheme can never hope to be a complete test of all possible deviations. Several such schemes have been proposed; for example, spacetimes that satisfy Einstein's equations but possess arbitrary and independent multipole moments \cite{BarackCutler}. Such spacetimes contain naked singularities, closed time-like curves and other non-physical matter distributions near the origin. In this paper another set of deformed Kerr metrics are used which retain the symmetries of the Kerr metric; namely stationarity, axisymmetry, reflection symmetry across the equatorial plane, and a second rank Killing tensor \cite{VYS}. The resultant bumpy BH metrics are described by a small number of dimensionless parameters. The symmetry of the bumpy BH metrics, in particular the existence of the Killing tensor, ensures that geodesics motion is regular and tri-periodic (in the $r$, $\theta$ and $\phi$ directions). This regularity is likely necessary for matched filtering to be able extract the waveform from the detector output. 

The bumpy BH metrics closely resemble Kerr at large radii, it therefore difficult to constrain these metrics using existing observations. The possibility of placing constraints using X-ray observations of fluorescent Iron (Fe) K$\alpha$ line emission, or broadband thermal emssion, from accretion disks was considered in \cite{PhysRevD.92.024039}; it was found that marginal constraints (i.e.\ constraining one of the small, dimensionless bump parameters to be less than unity) could generally only be achieved for the leading order deformations, and even then only when the BH was rapidly spinning. There has so far been no attempt to constrain deformations to the Kerr metric of the type considered here using GW observations.

The plan of this paper is as follows. In Sec.~\ref{BC} the formalism used for calculating the gravitational waveforms is described following the approach of \cite{BarackCutler2008}. In Sec.~\ref{GY} the deformed Kerr metrics first introduced in \cite{VYS} are described along with the modifications they induce to the Barack and Cutler waveforms which were first calculated in \cite{GairYunes}. The formalisms of signal analysis for extracting values of the system parameters and their associated errors from the waveform are then discussed in Sec.~\ref{sec:signal}. The results of this analysis are presented in Sec.~\ref{results} and finally discussions and concluding remarks are given in Sec.~\ref{sec:discussion}. Throughout this paper we use natural units, where c=G=1.

\section{EMRI Waveforms}\label{BC}

This section summarises the formalism presented by Barack \& Cutler \cite{BarackCutler2008} or calculating approximate ``analytic kludge'' waveforms from an EMRI around a Kerr BH. This formalism treats the binary at each instant as being purely Newtonian and emitting a Peters \& Mathews \cite{petersmathews1963} waveform. Post-Newtonian (PN) equations are then used to evolve the orbit through the inspiral. As the orbit is in the strong gravitational field, the analytic kludge is not accurate enough to produce EMRI template waveforms for detection (although augmented variants can approach the accuracy required for this purpose \cite{2017arXiv170504259C,CG2015}). However, the model is computationally efficient and its waveforms include several qualitative features of a true EMRI signal such as relativistic precession and radiation reaction. Hence it is ideal for use in provisional assessments of LISA's capability to perform source parameter estimation.

For an EMRI with component masses $\mu\ll M$, the inertia tensor is given by $I^{ij}(t)=\mu x^{i}(t)x^{j}(t)$, where $\mathbf{x}$ is the position of the CO relative to the central BH. Approximating the instantaneous motion of the CO as a Newtonian orbit with orbital frequency $\nu$, the second time derivative of $I^{ij}$ may be decomposed into $n$-harmonics of $\nu$ as $\ddot{I}^{ij}=\sum_{n}\ddot{I}_{n}^{ij}$. With a $z$-axis normal to the orbital plane, the three independent components of $\ddot{I}_{n}^{ij}$ are given by
\begin{equation}
\ddot{I}_{n}^{11}=a_{n}+c_{n},\quad\ddot{I}_{n}^{12}=b_{n},\quad\ddot{I}_{n}^{22}=c_{n}-a_{n},
\end{equation}
along with \cite{petersmathews1963}
\begin{align} 
a_{n}={}&-n{\cal{A}}\bigg[J_{n-2}(ne)-2eJ_{n-1}(ne)+\frac{2}{n}J_{n}(ne)+2eJ_{n+1}(ne)-J_{n+2}(ne)\bigg]\\ &\times\cos\left(n\Phi\right),\nonumber\\
b_{n}={}&-n{\cal{A}}\left(1-e^{2}\right)^{1/2}\left[J_{n-2}(ne)-2J_{n}(ne)+J_{n+2}(ne)\right]\sin\left(n\Phi\right),\\
c_{n}={}&2{\cal{A}}J_{n}(ne)\cos\left(n\Phi\right),\\
{\cal{A}}={}&\mu \left( 2\pi M\nu \right)^{2/3},
\end{align}
where $e$ is the orbital eccentricity, $\Phi(t)$ is the mean anomaly (such that $\dot{\Phi}=2\pi\nu$), and the $J_{n}$ denote Bessel functions of the first kind.

At the detector location, it is convenient to choose a coordinate frame such that the $z$-axis is aligned with the unit vector $\hat{\mathbf{r}}$ pointing from detector to source. The other axes are defined relative to the orbital angular momentum $\mathbf{L}$ of the binary, and are aligned with the basis vectors
\begin{equation}
\hat{\mathbf{p}}=\frac{\hat{\mathbf{r}}\times\hat{\mathbf{L}}}{\left|\hat{\mathbf{r}}\times\hat{\mathbf{L}}\right|},\quad\hat{\mathbf{q}}=\hat{\mathbf{p}}\times\hat{\mathbf{r}}.
\end{equation}
In the transverse--traceless gauge, the retarded metric perturbation at the detector due to a source at luminosity distance $D$ is then given (at leading quadrupole order) by \cite{MTW}
\begin{equation}
h_{ij}=\frac{2}{D}\left(P_{ik}P_{jl}-\frac{1}{2}P_{ij}P_{kl}\right)\ddot{I}^{kl},\quad h^{\pluscross}=\frac{1}{2}h^{ij}H_{ij}^{\pluscross},
\end{equation}
with the polarisation and transverse projection tensors
\begin{equation}
H_{ij}^{+}=\hat{p}_{i}\hat{p}_{j}-\hat{q}_{i}\hat{q}_{j},\quad H_{ij}^{\times}=\hat{p}_{i}\hat{q}_{j}+\hat{q}_{i}\hat{p}_{j},\quad P_{ij}=\delta_{ij}-\hat{r}_{i}\hat{r}_{j},
\end{equation}
where $\delta_{ij}$ is the Kronecker delta.

\begin{figure}[h!]
\begin{center}
\includegraphics[trim=0cm 0cm 0cm 0cm,width=0.65\textwidth]{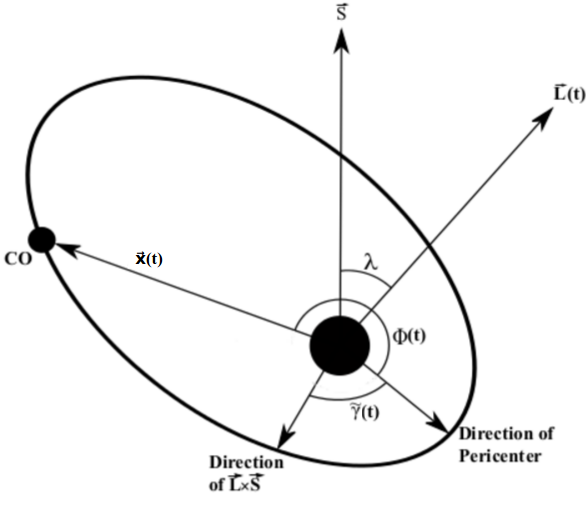}
\caption{The orbit of the CO about the central BH is modelled as an ellipse with evolving eccentricity $e(t)$. The orbital angular momentum $\mathbf{L}(t)$ of the system precesses about the spin angular momentum $\mathbf{S}$ of the central BH with angle $\alpha(t)$ (not shown). These two vectors define the inclination angle $\lambda$ and the precession angle $\tilde{\gamma}(t)$, which measures the intrinsic direction of pericentre with respect to $\mathbf{L}\times\mathbf{S}$. The position of the CO is parametrised by the mean anomaly $\Phi(t)$, measured from pericentre.\label{diagram}}
\end{center}
\end{figure}

The amplitudes of the two polarisation states may now be written in terms of the Peters--Mathews harmonic decomposition as \cite{BarackCutler2008}
\begin{align}
{}&h^{\pluscross}=\frac{1}{D}\sum_{n}A^{\pluscross}_{n}\,,\label{eq:PMharmonicsMOVED}\\
{}&A_{n}^{+}=\left[1+\left(\hat{\mathbf{r}}\cdot\hat{\mathbf{L}}\right)^{2}\right]\left[b_{n}\sin\left(2\gamma\right)-a_{n}\cos\left(2\gamma\right)\right]+\left[1-\left(\hat{\mathbf{r}}\cdot\hat{\mathbf{L}}\right)^{2}\right]c_{n},\label{eq:Anplus}\\
{}&A_{n}^{\times}=2\left(\hat{\mathbf{r}}\cdot\hat{\mathbf{L}}\right)\left[b_{n}\cos\left(2\gamma\right)+a_{n}\sin\left(2\gamma\right)\right].\label{eq:Ancross}
\end{align}
Here $\gamma$ is an azimuthal angle measuring the direction of pericentre with respect to the orthogonal projection of $\hat{\mathbf{r}}$ onto the orbital plane; it is further decomposed into an intrinsic part $\tilde{\gamma}$ that is used to parametrise the model (see Fig.~\ref{diagram} for visualisation), and an extrinsic part $\beta$ that may be written in terms of other extrinsic parameters (see \cite{BarackCutler2008} for explicit formulae). It is also convenient to express the orientation of $\hat{\mathbf{L}}$ with respect to 
the direction of the spin vector of the central BH, 
$\hat{\mathbf{S}}$, as an inclination angle $\lambda$ and an azimuthal angle $\alpha$ in the spin-equatorial plane (the latter is defined relative to a fixed ecliptic-based coordinate system \cite{PhysRevD.57.7089}). Writing the orientation of $\hat{\mathbf{S}}$ as $(\theta_{K},\phi_{K})$ in ecliptic coordinates, we have \cite{BarackCutler2008}
\begin{equation}\hat{\mathbf{L}}=\hat{\mathbf{S}}\cos\lambda+\frac{\hat{\mathbf{z}}-\hat{\mathbf{S}}\cos\theta_{K}}{\sin\theta_{K}}\sin\lambda\cos\alpha+\frac{\hat{\mathbf{S}}\times\hat{\mathbf{z}}}{\sin\theta_{K}}\sin\lambda\sin\alpha,\end{equation}
where $\hat{\mathbf{z}}=[0,0,1]^T$ is normal to the ecliptic plane.

With the above prescription for computing the instantaneous radiation from an extreme-mass-ratio Newtonian binary, relativistic effects are now added to the model by using PN expressions to evolve the relevant orbital parameters. The three phase angles $(\Phi,\tilde{\gamma},\alpha)$ are evolved with \cite{BarackCutler2008}
\begin{align}
\dot{\Phi}={}&2\pi\nu,\label{eq:Phi_dot}\\
\dot{\tilde{\gamma}}={}&6\pi\nu\left(2\pi M\nu\right)^{2/3}\left(1-e^{2}\right)^{-1}\left[1+\frac{1}{4}\left(2\pi M\nu\right)^{2/3}\left(1-e^{2}\right)^{-1}\left(26-15e^{2}\right)\right]\nonumber\\
&-12\pi\nu\cos\lambda\left(\frac{a}{M}\right)\left(2\pi M\nu\right)\left(1-e^{2}\right)^{-3/2},\label{eq:gamma_dot}\\
\dot{\alpha}={}&4\pi\nu\left(\frac{a}{M}\right)\left(2\pi M\nu\right)\left(1-e^{2}\right)^{-3/2},\label{eq:alpha_dot}
\end{align}
where $a=|\mathbf{S}|/M$ is the specific spin angular momentum. Eq.~\eqref{eq:Phi_dot} follows from the definition of the mean anomaly, while \eqref{eq:gamma_dot} and \eqref{eq:alpha_dot} introduce pericentre precession and Lense--Thirring precession. The orbital frequency and eccentricity are evolved with \cite{BarackCutler2008}
\begin{align}
\dot{\nu}={}&\frac{96}{10\pi}\left(\frac{\mu}{M^{3}}\right)\left(2\pi M\nu\right)^{11/3}\left(1-e^{2}\right)^{-9/2}\nonumber\\
&\left\{\left[1+\frac{73e^{2}}{24}+\frac{37e^{4}}{96}\right]\left(1-e^{2}\right)\right.+\left(2\pi M\nu\right)^{2/3}\left[\frac{1273}{336}-\frac{2561e^{2}}{224}-\frac{3885e^{4}}{128}-\frac{13147e^{6}}{5376}\right]\nonumber\\
&-\left(2\pi M\nu\right)\left(\frac{a}{M}\right)\cos\lambda\left(1-e^{2}\right)^{-1/2}\left[\frac{73}{12}+\frac{1211e^{2}}{24}\left.+\frac{3143e^{4}}{96}+\frac{65e^{6}}{64}\right]\right\},\label{eq:nu_dot}\\
\dot{e}={}&-\frac{e}{15}\left(\frac{\mu}{M^{2}}\right)\left(1-e^{2}\right)^{-7/2}\left(2\pi M\nu\right)^{8/3}\left[\left(304+121e^{2}\right)\left(1-e^{2}\right)\left(1+12\left(2\pi M\nu\right)^{2/3}\right)\right.\nonumber\\
&\left.-\frac{1}{56}\left(2\pi M\nu\right)^{2/3}\left(8\times16705+12\times9082e^{2}-25211e^{4}\right)\right]\nonumber\\
&+e\left(\frac{\mu}{M^{2}}\right)\left(\frac{a}{M}\right)\cos\lambda \left(2\pi M\nu\right)^{11/3}\left(1-e^{2}\right)^{-4}\left[\frac{1364}{5}+\frac{5032e^{2}}{15}+\frac{263e^{4}}{10}\right],\label{eq:e_dot}
\end{align}
which introduce inspiralling and circularisation respectively. The inclination angle also evolves due to radiation reaction in a fully relativistic treatment, but very slowly \cite{hughes2000evolution}; hence $\lambda$ is approximated as constant in the analytic kludge \cite{BarackCutler2008}.

It is known that the angular rates $(\dot{\Phi},\dot{\tilde{\gamma}},\dot{\alpha})$ do not agree in general with the corresponding values $(\omega_r,\omega_\theta-\omega_r,\omega_\phi-\omega_\theta)$ for a Kerr EMRI \cite{CG2015}, where $\omega_{\{r,\theta,\phi\}}$ are the fundamental frequencies of radial, polar and azimuthal motion on a Kerr geodesic \cite{schmidt2002celestial}. In this model, we correct the angular rates at the start of evolution (but not along the inspiral), using a parameter-space map $(M,a,\nu)\mapsto(M',a',\nu')$ such that \cite{2017arXiv170504259C}
\begin{align}
\dot{\Phi}(M',a',\nu')={}&\omega_r(M,a,\nu),\\
\dot{\tilde{\gamma}}(M',a',\nu')={}&\omega_\theta(M,a,\nu)-\omega_r(M,a,\nu),\\
\dot{\alpha}(M',a',\nu')={}&\omega_\phi(M,a,\nu)-\omega_\theta(M,a,\nu).
\end{align}
While this correction does not address the accumulated phase error of the analytic kludge (with respect to more accurate EMRI models) over the full inspiral, it is computationally negligible and yields more physically accurate waveforms over short timescales.

If the spin of the CO is neglected, an EMRI event is completely specified by 14 degrees of freedom. With the choice of some arbitrary reference frequency $\nu_{0}$ in the detector's sensitivity band, the event time $t_{0}$ is defined as the instant the orbital frequency equals $\nu_{0}$. (Hereafter, a subscript $0$ indicates the value a time-dependent quantity takes at time $t_{0}$, e.g. $\alpha_{0}=\alpha(t=t_{0})$.) We choose 14 dimensionless quantities to parametrise the EMRI model:
\begin{align}\label{eq:sys_par}
&\left\{\log_{10}\left(\frac{\mu}{M_\odot}\right),\log_{10}\left(\frac{M}{M_\odot}\right),\frac{a}{M},e_{0},\cos(\lambda),\tilde{\gamma}_{0},\Phi_{0}, \right.\\ &\quad\quad\left. \cos(\theta_{S}),\phi_{S},\cos(\theta_{K}),\phi_{K},\alpha_{0},\log_{10}\left(\frac{D}{\mathrm{Gpc}}\right),\frac{t_{0}}{M}\right\},\nonumber
\end{align}
where all logarithms are decadic and $(\theta_{S},\phi_{S})$ is the sky position of the source (i.e. the orientation of $\hat{\mathbf{r}}$ in ecliptic coordinates). The first seven parameters are \emph{intrinsic} to the source itself, while the remaining seven depend on its position and orientation relative to the ecliptic plane.

The final piece of the waveform model is the computation of a detector's reponse to the astrophysical signal $h^{\pluscross}$, which enables the model to be used in data analysis. The three arms of a LISA-like detector function as two Michelson interferometers, from which two independent strain signals $h_{\{I,II\}}$ may be obtained. These admit the same harmonic decomposition as $h^{\pluscross}$, and are related to \eqref{eq:Anplus} and \eqref{eq:Ancross} by \cite{PhysRevD.57.7089}
\begin{equation}
h_{\{I,II\}}=\sum_n\frac{1}{D}\frac{\sqrt{3}}{2}\left(F_{\{I,II\}}^{+}A_{n}^{+}+F_{\{I,II\}}^{\times}A_{n}^{\times}\right),
\end{equation}
where the antenna pattern functions \cite{apostolatos1994spin}
\begin{align}
F_{I}^{+}={}&\frac{1}{2}\left(1+\cos^{2}\theta\right)\cos\left(2\phi\right)\cos\left(2\psi\right)-\cos\theta\sin\left(2\phi\right)\sin\left(2\psi\right),\\
F_{I}^{\times}={}&\frac{1}{2}\left(1+\cos^{2}\theta\right)\cos\left(2\phi\right)\sin\left(2\psi\right)+\cos\theta\sin\left(2\phi\right)\cos\left(2\psi\right),\\
F_{II}^{+}={}&\frac{1}{2}\left(1+\cos^{2}\theta\right)\sin\left(2\phi\right)\cos\left(2\psi\right)+\cos\theta\cos\left(2\phi\right)\sin\left(2\psi\right),\\
F_{II}^{\times}={}&\frac{1}{2}\left(1+\cos^{2}\theta\right)\sin\left(2\phi\right)\sin\left(2\psi\right)-\cos\theta\cos\left(2\phi\right)\cos\left(2\psi\right),
\end{align}
depend on the sky location $(\theta,\phi)$ and the polarisation angle $\psi$ of the source in a precessing detector-based coordinate system (see \cite{BarackCutler2008} for formulae relating $(\theta,\phi,\psi)$ to $(\theta_{S},\phi_{S},\theta_{K},\phi_{K})$ in ecliptic coordinates). Finally, Doppler modulation of the waveform phase is included through the map \cite{BarackCutler2008}
\begin{equation}
\Phi\mapsto\Phi+2\pi\nu R\sin\theta_{S}\cos\left(\frac{2\pi t}{T}-\phi_{S}\right),
\end{equation}
where $R=1\,\mathrm{AU}$ and $T=1\,\mathrm{yr}$. This map accounts for the orbital motion of LISA, but neglects the smaller effects of the detector's cartwheeling motion \cite{CornishRubbo}.

\section{EMRIs Around Bumpy BHs}\label{GY}
This goal of this paper is to use EMRI gravitational waveforms to constrain deviations from the Kerr metric. In Sec.\ref{sec:metrics} the family of parametrically deformed Kerr metrics first introduced by \cite{VYS} is described and Sec.\ref{sec:deltadot} summarises the results of \cite{GairYunes} which allow the EMRI waveform model described above to be generalised to these new metrics.

\subsection{Bumpy Black Hole Spacetimes}\label{sec:metrics}
The metrics described in this section are all continuously parameterised smooth deformations of the Kerr metric which retain the properties of stationarity, axisymmetry and the existence of a second rank Killing tensor (at least to leading order in the metric deformation parameter). These symmetries ensure that geodesic motion in these deformed metrics will continue to possess four constants of motion: these constants can be chosen to be the test particle rest mass, the energy, the $z$-component of angular momentum and a fourth constant which smoothly recovers the Carter constant in the limit that the deformation from the Kerr metric tends to zero. The deformed metrics, known as ``bumpy black holes'', are not required to satisfy Einstein's equations.

Bound orbits around a BH with the symmetries described are characterised by three fundamental frequencies which can be associated with motion in the radial, azimuthal and polar directions. The GWs from a test particle on a bound orbit contains a superposition of harmonics of these three fundamental frequencies. The metric deformation changes the three fundamental frequencies and hence the spectral content of the GWs. It should be noted that only the metric tensor is varied from the standard GR case; it would also be possible to consider modifications to the sourcing of GWs by the CO, the propagation or polarisation content of the GWs, or the radiation back-reaction on the CO. Such additional changes are not considered here.
 
The starting point for deriving these metrics is the stationary, axisymmetric Lewis-Papapetrou metric in $(t,\rho,\phi,z)$ coordinates with line element 
\begin{equation} \label{eq:LewisPapapetrouMetric} \textrm{d}s^{2}=-V\left(\textrm{d}t-\frac{q}{V} \textrm{d}\phi\right)^{2}+\frac{\rho^{2}}{V}\textrm{d}\phi ^{2}+\gamma\,\textrm{d}\rho^{2}+\lambda\,\textrm{d} z^{2} \; , \end{equation}
where $V,q,\gamma$ and $\lambda$ are functions of $\rho$ and $z$. This metric has both timelike (${t^{\mu}\!=\!\{1,0,0,0\}}$) and axial (${l^{\mu}\!=\!\{0,0,0,1\}}$) Killing vector fields satisfying
\begin{equation} \nabla_{(\mu}t_{\nu )}=\nabla_{(\mu}l_{\nu )}=0\; . \end{equation}
The metric in Eq.\ref{eq:LewisPapapetrouMetric} can be transformed into Boyer-Lindquist-like coordinates $(t,r,\theta,\phi)$ where $r$ and $\theta$ are defined implicitly by $\rho = \sqrt{\Delta}\sin\theta$ and $z=(r-M)\sin\theta$ (with $\Delta=r^{2}-2r+a^{2}$). Expanding the functions $V,q,\gamma$ and $\lambda$ as the Kerr expressions (with dimensionless spin parameter $a$) plus a perturbation (e.g.\ $V=\bar{V}+\epsilon\,\delta V$) it follows that
\begin{equation}\label{eq:introduceepsilon} \textrm{d}s^{2}=\bar{g}_{\mu\nu}\textrm{d}x^{\mu}\textrm{d}x^{\nu}+\epsilon\, h_{\mu\nu}\textrm{d}x^{\mu}\textrm{d}x^{\nu}\;. \end{equation}
Here $\bar{g}_{\mu\nu}$ denotes the standard Kerr metric, $\epsilon$ is a bookkeeping parameter, and $h_{\mu\nu}$ has the following non-zero Boyer-Lindquist-like coordinate components:
\begin{eqnarray}
h_{tt} &= - \delta V \;,\quad
h_{t \phi} = \delta q \;, \quad
h_{rr} = \delta \lambda \cos^{2}{\theta} + \delta\gamma \,\frac{(r-M)^2\sin^2\theta}{\Delta} \;,\nonumber \\
h_{r\theta} &= \left(r-M\right) \cos{\theta} \sin{\theta} \left( \delta \gamma - \delta \lambda \right)\;, \nonumber \\h_{\theta \theta} & =\delta \lambda \sin^{2}{\theta} \left(r-M\right)^{2} + \delta \gamma \Delta\cos^{2}\theta\;,\nonumber\\
h_{\phi \phi} &= \frac{\sin^2\theta}{\rho^{2}-2Mr} \left\{ 4aMr \, \delta q - \left[ \left(r^2+a^2\right)^2 - a^2\sin^2\theta \Delta \right] \, \delta V \right\} \; .
\end{eqnarray}
The remaining freedom in the metric is constrained, not by requiring Einstein's equations to be satisfied, but instead by requiring that there exists an (approximate) second rank Killing tensor, $\xi_{\mu\nu}$, satisfying
\begin{equation} \nabla_{(\lambda}\xi_{\mu\nu )} = \mathcal{O}(\epsilon^{2})\;. \end{equation}
In addition to requiring the existence of a Killing tensor, the remaining freedom in the metric is further reduced by requiring that $h_{\mu\nu}$ tends to zero at spatial infinity faster that $(r/M)^{-2}$, which ensures the perturbed metric remains asymptotically flat with the same mass and spin as the background Kerr metric.

These constraints force the $h_{r\theta}$ and $h_{\theta\theta}$ components to vanish. The remaining non-zero components can be expanded in the ``weak field'' limit, 
\begin{equation} h_{\mu\nu}=\sum_{n}h_{\mu\nu,n}\left(\frac{M}{r}\right)^{n} \;. \end{equation}
Expressions for the $h_{\mu\nu,n}$ coefficients up to and including $n=5$ were derived in \cite{GairYunes}.
 
Here it is convenient to introduce the ${\cal{B}}_{N}$ notation of \cite{GairYunes}. The general solutions described in \cite{GairYunes} contain a number of unknown functions of radius, labelled $\gamma_m(r)$, which are expanded in powers of $1/r$; the dimensionless constant $\gamma_{m,n}$ is the coefficient of $1/r^n$ term in the expansion of $\gamma_m(r)$. 
If only the leading ${\cal{O}}(M^{2}/r^{2})$ terms are retained then the metric deformation is completely specified by four dimensionless constants ${\cal{B}}_{2}\!=\!\left\{ \gamma_{1,2},\gamma_{3,1},\gamma_{3,3},\gamma_{4,2} \right\}$. If instead terms up to ${\cal{O}}(M^{3}/r^{3})$ are retained then the deformation is specified by the seven constants ${\cal{B}}_{2}\cup{\cal{B}}_{3}$ with ${\cal{B}}_{3}\!=\!\left\{ \gamma_{1,3},\gamma_{3,4},\gamma_{4,3} \right\}$. Retaining terms up to ${\cal{O}}(M^{4}/r^{4})$ means that he deformation is specified by the ten constants ${\cal{B}}_{2}\cup{\cal{B}}_{3}\cup{\cal{B}}_{4}$ with ${\cal{B}}_{4}\!=\!\left\{ \gamma_{1,4},\gamma_{3,5},\gamma_{4,4} \right\}$. Retaining terms up to ${\cal{O}}(M^{5}/r^{5})$ means that the deformation is specified by the thirteen constants ${\cal{B}}_{2}\cup{\cal{B}}_{3}\cup{\cal{B}}_{4}\cup{\cal{B}}_{5}$ with ${\cal{B}}_{5}\!=\!\left\{ \gamma_{1,5},\gamma_{4,5},\gamma_{3,6} \right\}$. In this paper terms of higher orders will not be considered. In addition we will set $\gamma_{3,1}=0$ in ${\cal B}_2$ so there are a total of twelve constants, three in each of the four sets. The restriction $\gamma_{3,1}=0$ was originally made in~\cite{GairYunes} to ensure that the inclination of the orbit remained constant under radiation reaction in the weak field limit. For consistency with that work we make the same choice here. For the rest of this paper, when referring to the ${\cal{B}}_{N}$ limit we will mean that all the constants $\gamma_{m,n}\!=\!0$ except for those quantities in the set ${\cal{B}}_{N}$.
 
The perturbed, or ``bumpy'', black holes described here represent an agnostic approach to parameterising possible deviations from the Kerr metric in the sense that no particular underlying theory of gravity has been assumed. However, it should be noted that known BH solutions in some specific alternative theories may be recovered within this framework by making specific choices for the constants $\gamma_{m,n}$. For example, the slowly rotating (i.e. linear in spin, $a$) BH solution to dynamical Chern-Simons (dCS) gravity \cite{PhysRevD.68.104012, PhysRevD.79.084043} gravity is obtained by setting all the constants $\gamma_{m,n}\!=\!0$ except for 
\begin{eqnarray} 
&\gamma_{3,5}=-\frac{3}{5}a\zeta \; , \quad \gamma_{3,6}=-\frac{65}{28}a\zeta\; , \quad \gamma_{3,7}=-\frac{709}{112}a\zeta \; , \nonumber\\
\Rightarrow\;&\textrm{d}s_{\textrm{dCS}}^{2}=\bar{g}_{\mu\nu}\textrm{d}x^{\mu}\textrm{d}x^{\nu}+\frac{5}{8}\zeta\frac{aM^{5}}{r^{4}}\left(1+\frac{12M}{7r}+\frac{27M^{2}}{10r^{2}}\right)\sin^{2}\theta \; . \label{eq:dCSmetric} \end{eqnarray}
A quadratic in spin solution to dCS is also known \cite{PhysRevD.86.044037}, however this cannot be reproduced exactly within the current framework as it does not possess a second rank Killing tensor.

\subsection{Modified EMRI Waveforms}\label{sec:deltadot}

The EMRI waveform model described in Sec.~\ref{BC} is based on elliptical Keplerian orbits. These ellipses are forced to precess, mimicking the more complicated geodesic orbits in the Kerr metric; the precession frequencies are calculated from the frequencies of Kerr geodesics. The EMRI waveform model can be extended to the bumpy BH metrics described in Sec.~\ref{GY} by replacing the Kerr precession frequencies with frequencies calculated from geodesics in the perturbed metrics. Changes to the precession frequencies also lead, via the quadrupole formula, to changes in the inspiral rate; corrections to all of the evolution equations in Eqs.~\ref{eq:Phi_dot} to \ref{eq:e_dot} were calculated in \cite{GairYunes}.

In order to calculate the appropriate precession frequencies it is first necessary to identify each geodesic in Kerr with a corresponding geodesic in the bumpy BH metric. This is achieved by requiring the orbit to have the same shape, i.e. the turning points in the $r,\,\theta,\,\phi$ motion occur at the same Boyer-Linquist coordinate locations. Provided we consider only bound orbits this give a suitable $1\textrm{--}1$ map. Using this map \cite{GairYunes} calculated how the perturbation to the constants of motion (${E=t_{\mu}u^{\mu},\,L_{z}=l_{\mu}u^{\mu},\,\textrm{and }Q=\xi_{\mu\nu}u^{\mu}u^{\nu}}$, where $u^{\mu}$ is the CO four-velocity) induced by the metric perturbation affects the three orbital frequencies $\Omega_{r}$, $\Omega_{\theta}$ and $\Omega_{\phi}$. The leading order ${\cal{B}}_{2}$ corrections to the evolution equations in Eqs.~\ref{eq:Phi_dot} to \ref{eq:e_dot} are given in Eqs.~\ref{deltadot} to \ref{deltadotend}.
\begin{align} \label{deltadot}
 2\pi M^{2} \;\delta \left( \frac{d\nu}{dt} \right)_{{\cal{B}}_{2}} &= \frac{16}{5}\eta\frac{\left( 2\pi M\nu \right)^{13/3}}{\left( 1-e^{2} \right)^{9/2}} \left( 18+78e^{2}+\frac{99}{4}e^{4} \right) \left( \gamma_{1,2}+2\gamma_{4,2} \right) \\
 M\;\delta \left(\frac{d\gamma}{dt}\right)_{{\cal{B}}_{2}} &= \frac{\left( 2\pi M\nu \right)^{5,3}}{2\left(1-e^{2}\right)}\left( \gamma_{1,2}+2\gamma_{4,2} \right) \\
 \delta \left(\frac{de}{dt}\right)_{{\cal{B}}_{2}} &=-\frac{16}{5}\eta\frac{\left(2\pi M\nu\right)^{10/3}}{\left(1-e^{2}\right)^{7/2}}\left( \frac{93}{4}e +\frac{67}{4}e^{3} + \frac{1}{4}e^{5} \right)\left( \gamma_{1,2}+2\gamma_{4,2} \right)  \\
 M\;\delta \left(\frac{d\alpha}{dt}\right)_{{\cal{B}}_{2}} &= -\frac{a\left( 2\pi M \nu \right)^{2}}{\left( 1-e^{2} \right)^{3/2}} \left( \gamma_{1,2}+2\gamma_{4,2} \right) \label{deltadotend}\end{align}
 These expressions are taken from Eqs.~(327)--(330) in~\cite{GairYunes}, which include the restriction $\gamma_{3,1}=0$ mentioned earlier.
The corresponding expression for the ${\cal{B}}_{3}$ to ${\cal{B}}_{5}$ metric perturbations are given by Eqs.~\ref{deltadotn} to \ref{deltadotendn}, and are taken from Eqs.~(331)--(334) in~\cite{GairYunes}. (Explicit expressions for the eccentricity dependent factors ${g_{\nu,N},\,g_{\gamma,N},\,g_{e,N},\,\textrm{and}\;g_{\alpha,N}}$ are given in Eqs.~(301)--(302) and (335)--(346)) of~\cite{GairYunes}.)
\begin{align}\label{deltadotn}
2\pi M^{2} \; \delta\left( \frac{d\nu}{dt} \right)_{{\cal{B}}_{N}}&=\frac{16}{5}\eta\frac{\left(2\pi M\nu\right)^{2N/3+2}}{\left(1-e^{2}\right)^{N+5/2}}g_{\nu , N}(e)\left(\gamma_{1,N}+2\gamma_{4,N}\right) \\
M\delta \; \left(\frac{d\gamma}{dt}\right)_{{\cal{B}}_{N}}&=\frac{\left(2\pi M\nu\right)^{(2N+1)/3}}{\left(1-e^{2}\right)^{N-1}}g_{\gamma , N}(e) \left(\gamma_{1,N}+2\gamma_{4,N}\right) \\
 M \; \delta \left(\frac{de}{dt}\right)_{{\cal{B}}_{N}}&=\frac{-16}{5}\eta\frac{\left(2\pi M\nu\right)^{2N/3+2}}{\left(1-e^{2}\right)^{N+3/2}}g_{e,N}(e)\left(\gamma_{1,N}+2\gamma_{4,N}\right) \\
M\; \delta\left(\frac{d\alpha}{dt}\right)_{{\cal{B}}_{N}}&=-S\frac{\left(2\pi M\nu\right)^{2(N+1)/3}}{\left(1-e^{2}\right)^{N-1/2}}g_{\alpha,N}(e) \left(\gamma_{1,N}+2\gamma_{4,N}\right) \; .\label{deltadotendn}
\end{align}

\begin{figure}[h!]\label{waveform}
\begin{center}
\includegraphics[trim=1cm 0cm 1cm 0cm,width=0.75\textwidth]{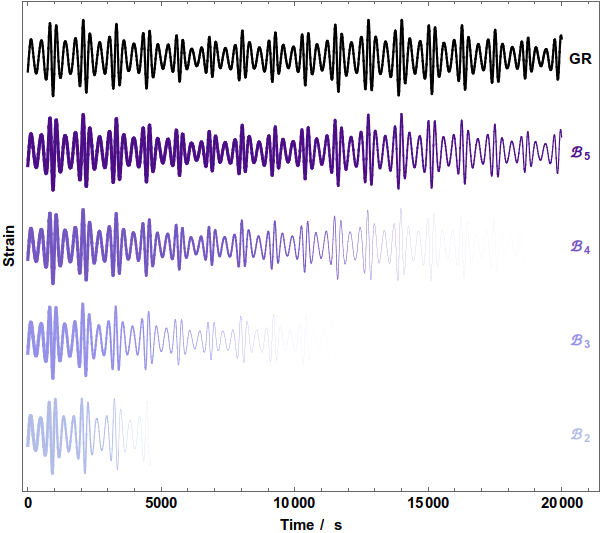}
\caption{Shown in black is a short section of the plus component of a gravitational waveform from an EMRI around a Kerr black hole approximately one year before plunge. The central BH has mass $M=10^{6}\Msun$ and spin $S=0.7$. The CO has mass $\mu=10\Msun$ and is on an orbit with inclination $\lambda=\pi/4$, semi-major axis $a=7M$ and eccentricity $e_{0}=0.3$ at time $t=0$. The other system parameters were set to ${\theta_{K}\!=\!\pi/8,\, \phi_{K}\!=\!0, \,\theta_{S}\!=\!\pi/4,\, \phi_{s}\!=\!\pi /2,\, \gamma_{0}\!=\!0, \,\Phi_{0}\!=\!0,\,\alpha_{0}\!=\!0}$. The precession effects of the orbital eccentricity and inclination are clearly imprinted on the EMRI waveform. The coloured curves show EMRI waveforms with the same system parameters but where the central BH is not Kerr; plots for the $\mathcal{B}_{N}$ (for $N\!=\!2,\,3,\,4,\,5$) metrics with $\epsilon=0.4$ are shown. The perturbed EMRIs waveforms are, by construction, in phase agreement with the Kerr waveform at $t=0$ but they gradually drift out of phase (the coloured curves are weighted by the match between the perturbed signal and the Kerr signal to highlight this dephasing). The $\mathcal{B}_{5}$ EMRI stays in phase longest as the metric perturbation is suppressed by a high power of $(M/r)<1$. \label{fig.examplewave}}
\end{center}
\end{figure}

It should be noted that these corrections depend upon a consistent combination of the small, dimensionless bump parameters; at the ${\cal{B}}_{N}$ order the combination is $\gamma_{1,N}+2\gamma_{4,N}$. Hereafter we let $\epsilon = \gamma_{1,N}+2\gamma_{4,N}$, and $\epsilon$ is treated as an additional free parameter in the model to be measured from the data. 

Adding the corrections in Eqs.~\ref{deltadot} to \ref{deltadotend} to the orbital evolution equations for the EMRI model in Eqs.~\ref{eq:Phi_dot} to \ref{eq:e_dot} allows the EMRI model to be extended to any of the bumpy BH spacetimes described in Sec.~\ref{GY}. Shown in Fig.~\ref{fig.examplewave} is an example EMRI waveform from the standard model around a Kerr BH, and example waveforms from the extended model around several bumpy BHs. Only a short segment (around $20000\,\textrm{s}$) of the waveform is shown, in reality it will be possible to observe the evolution of this system for several years (a few times $10^{5}$ orbits) before plunge. The EMRI waveforms in Fig.~\ref{fig.examplewave} have been generated with the same system parameters and are, by construction, in phase agreement at the start of the signal. The lines for the EMRI waveforms from bumpy BHs have been faded to illustrate how the signals rapidly drift out of phase with the standard Kerr signal; the higher order $\mathcal{B}_{N}$ signals dephase more slowly because the metric perturbation is suppressed by a higher power of $(M/r)$.

\section{Fundamentals of Signal Analysis and the Fisher Matrix}\label{sec:signal}
The LISA constellation effectively functions as two crossed Michelson detectors \cite{PhysRevD.57.7089} which are here labelled by the subscript index $\alpha\in\{I,II\}$. The output from these detectors is denoted $s_{\alpha}(t)$, and the following Fourier transform conventions are used;
\begin{equation} \tilde{s}_{\alpha}(f)=\int_{-\infty}^{+\infty}\textrm{d}t\;s_{\alpha}(t)e^{i2\pi f t}\;,\quad\textrm{and }\;s_{\alpha}(t)=\int_{-\infty}^{+\infty}\textrm{d}f\;\tilde{s}_{\alpha}(f)e^{-i2\pi f t}\;. \end{equation}
The measured signal in each detector is the sum of the instrumental noise and, possibly, a GW EMRI signal;  
\begin{equation} s_{\alpha}(t)=n_{\alpha}(t)+h_{\alpha}(t;\vec{\theta}_{0})\;, \end{equation}
where $\vec{\theta}_{0}$ is a vector of parameters describing the EMRI source. The instrumental noise is assumed to be zero-mean, stationary, Gaussian and uncorrelated between the two channels with (one-sided) power spectral density $S_{n}(f)$\footnote{The noise PSD was assumed to be that of the ``Classic LISA'' N2A5M5L6 mission described in \cite{2016PhRvD..93b4003K}. This is similar to the more recent noise curve described in \cite{2017arXiv170200786A} and produces nearly identical results when signals are normalised to a fixed SNR.}. Under these standard assumptions the noise is fully characterised by the following two--point expectation value
\begin{equation}\label{eq:noise}\left< \tilde{n}_{\alpha}(f)\tilde{n}_{\beta}(f ')\right> =\frac{1}{2}\delta(f-f ')S_{n}(f)\delta_{\alpha\beta}\;.\end{equation}
 
The \emph{likelihood} is the probability of obtaining the observed data given a particular value of $\vec{\theta}$ and its logarithm is given by 
\begin{equation} \label{eq:like} \log\left(\mathcal{L}(\vec{\theta})\right) = \frac{-\left(s_{\alpha}(t)-h_{\alpha}(t;\vec{\theta})|s_{\alpha}(t)-h_{\alpha}(t;\vec{\theta})\right)}{2} + \textrm{normalisation constant} \;, \end{equation}
where the following definition of the signal inner product has been used,
\begin{equation}\label{inner} \left( a_{\alpha}(t),b_{\alpha}(t) \right) = 4\Re \left\{\sum_{\alpha}\int_{0}^{\infty}\frac{\tilde{a}_{\alpha}^{*}(f)\tilde{b}_{\alpha}(f)}{S_{n}(f)}df \right\} \;. \end{equation}
The signal-to-noise-ratio (SNR) of the source is defined as $\varrho=\left(h_{\alpha}(t)|h_{\alpha}(t)\right)^{1/2}$. In the limit of large SNR the log--likelihood in Eq.~\ref{eq:like} may be expanded to quadratic order in $\delta\theta^{a}\equiv\theta^{a}-\theta_{0}^{a}$, giving
\begin{equation} \label{eq:likeexpand}\log\left(\mathcal{L}(\vec{\theta})\right) = \frac{-\left(n|n\right)-\delta\theta^{a}\delta\theta^{b}\Gamma_{ab}+2\delta\theta^{a}\left(n|\partial_{a}h\right)}{2} + \textrm{normalisation constant}\;, \end{equation}
where $\Gamma_{ab}=(\partial_{a}h|\partial_{b}h)$ and $\partial_{a}h=\partial h(t;\vec{\theta})/\partial\theta^{a}|_{\vec{\theta}=\vec{\theta}_{0}}$. This quadratic form in $\delta\vec{\theta}$ has a maximum at 
\begin{equation} \delta\theta^{a}_{\textrm{max}}=\Gamma^{ab}\left(n|\partial_{b}h\right) \;, \end{equation}
where $\Gamma^{ab}$ is the inverse of $\Gamma_{ab}$. The \emph{maximum likelihood estimator} for the source parameters, $\vec{\theta}_{\textrm{max}}\equiv\vec{\theta}_{0}+\delta\vec{\theta}_{\textrm{max}}$, is linear in $n$ and is therefore a Gaussian random variable with mean and covariance which can be calculated from Eq.~\ref{eq:noise};
\begin{equation} \textrm{mean}\left(\theta^{a}_{\textrm{max}}\right)=\theta^{a}_{0}\;,\quad \textrm{cov}\left(\theta^{a}_{\textrm{max}}\right) = \Gamma^{ab} \;. \end{equation}
The matrix $\Gamma_{ab}$ is known as the \emph{Fisher information matrix}, and the inverse $\Gamma^{ab}$ describes the covariance of the maximum likelihood estimator for the true source parameters in the limit of large SNR.

Evaluating the Fisher matix involves computing multiple signal inner products which are defined in terms of the Fourier transform of the EMRI waveform. For computational speed, and following \cite{BarackCutler2008}, an approximation to the inner product which can be evaluated in the time domain was used. First, define the \emph{noise weighted} waveform polarisation components (c.f.\ Eq.~\ref{eq:PMharmonicsMOVED})
\begin{equation} \hat{h}^{\small\{+,\times\small\}}(t)=\frac{1}{D}\sum_{n}\frac{A^{\small\{+,\times\small\}}_{n}(t)}{S_{n}^{1/2}\left(f_{n}(t)\right)} \,, \end{equation}
where $f_{n}$ is a combination of the radial and azimuthal orbital frequencies, 
\begin{equation} f_{n}(t)=n\nu(t)+\frac{\dot{\tilde{\gamma}}(t)}{\pi} \,. \end{equation}
The inner product can then be approximated as an integral of the \emph{noise weighted} waveforms in the time domain;
\begin{equation} \left( a(t),b(t) \right) \approx 2\sum_{\alpha}\int_{0}^{T}dt \; \hat{a}_{\alpha}(t)\hat{b}_{\alpha}(t) \,. \end{equation}
This approximation is exact for circular, equatorial orbits in the extreme mass ratio limit.

\begin{figure*}[h]
\begin{center}
    \includegraphics[width=0.7\textwidth]{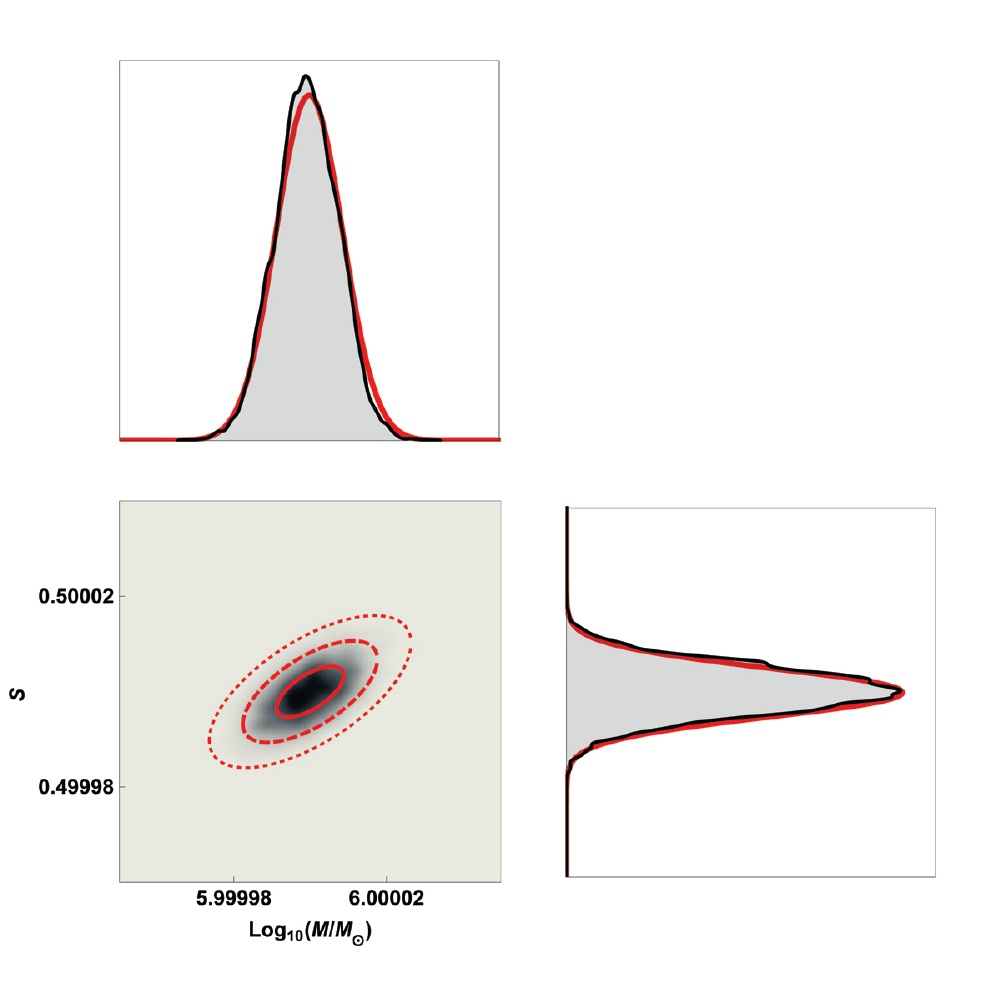}
  \caption{Marginalised posterior distributions, overlayed with Fisher matrix predictions, for the parameters describing the central BH (mass and spin). The shaded gray regions show the numerical results obtained using PolyChord, while the smooth red curves show the Fisher matrix predictions (in the 2 dimensional posterior 1, 2 and 3$\sigma$ contours are shown; the injected signal had an SNR of 30). \label{MCMCres_centralBH}
}
\end{center}
\end{figure*}

\begin{figure*}[h]
\begin{center}
    \includegraphics[width=0.7\textwidth]{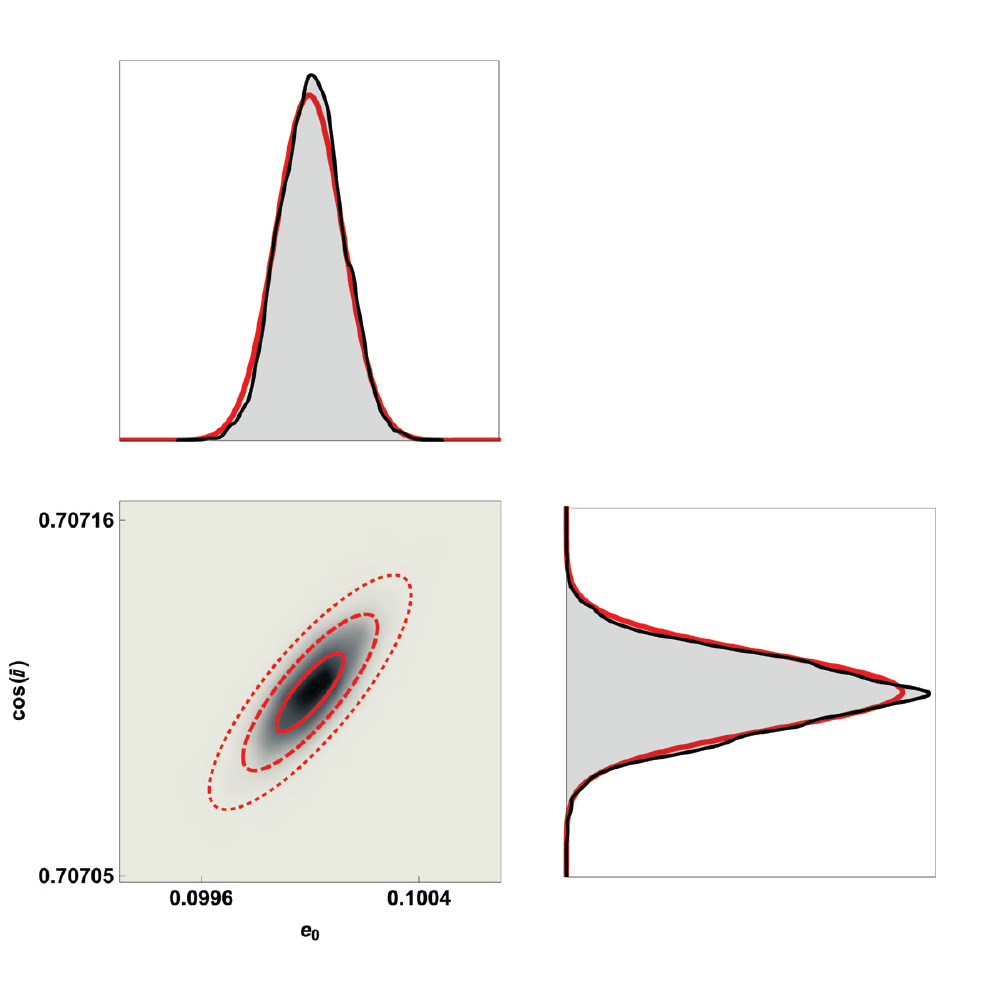}
  \caption{Marginalised posterior distributions, overlayed with Fisher matrix predictions, for the parameters describing the CO orbit (eccentricity and orbital inclination at the beginning of the observation). The shaded gray regions show the numerical results obtained using PolyChord, while the smooth red curves show the Fisher matrix predictions (in the 2 dimensional posterior 1, 2 and 3$\sigma$ contours are shown; the injected signal had an SNR of 30).
\label{MCMCres_orbit}
}
\end{center}
\end{figure*}

\begin{figure*}[h]
\begin{center}
    \includegraphics[width=0.38\textwidth]{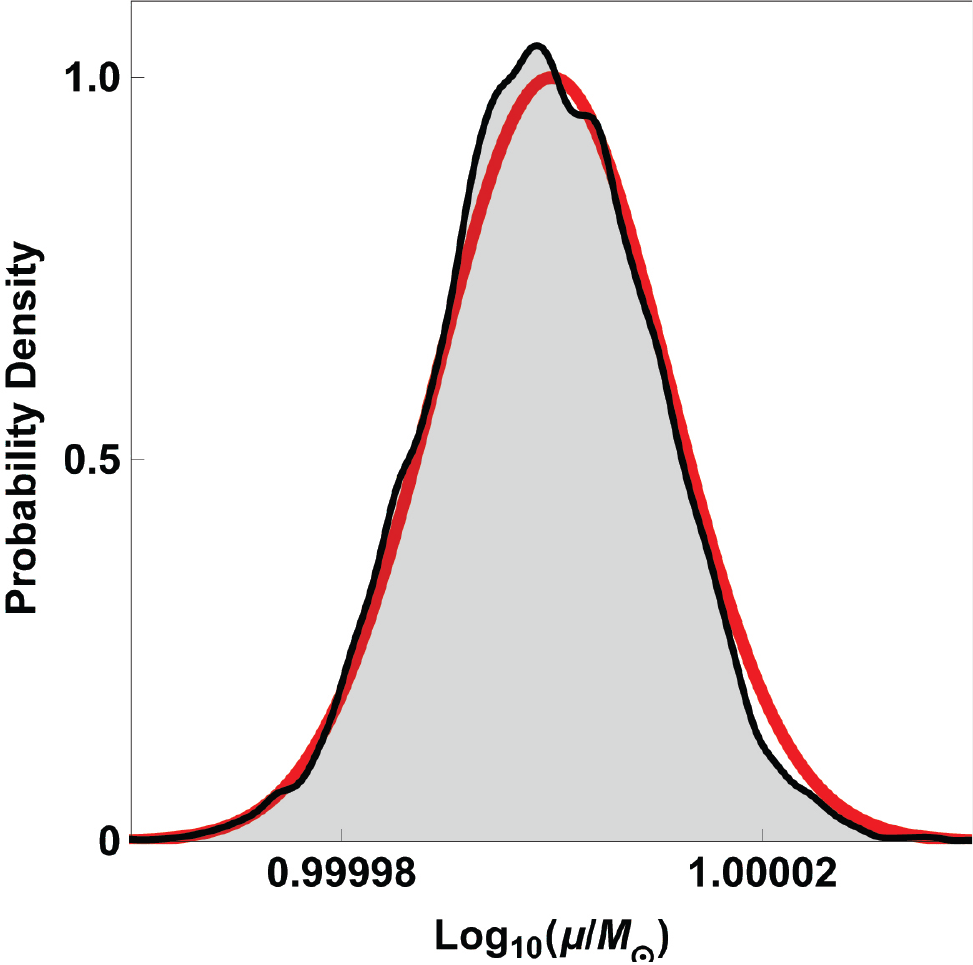}
  \caption{One--dimensional marginalised posterior distribution, overlayed with the Fisher matrix prediction, for the CO mass. The shaded gray region shows the numerical results obtained using PolyChord, while the red curve shows the Fisher matrix prediction (the injected signal had an SNR of 30). \label{MCMCres_SmallBH}
}
\end{center}
\end{figure*}

\subsection{Verifiying the Fisher Matrix Results}\label{sec:verify}
The Fisher matrix provides a computationally efficient method to estimate the precision with which the source parameters can be measured. It may be evaluated relatively quickly, and hence is well suited to exploring how parameter estimation precision varies across a large, high-dimensional parameter space. However, it is well known in the context of GW parameter estimation that the Fisher matrix must be used with caution (see, e.g.\ \cite{2013PhRvD..88h4013R, PhysRevLett.107.191104}); in some cases, it has been shown to misestimate the true uncertainty by several orders of magnitude (Fisher matrices will certainly not be used for real LISA data analysis). For EMRIs, the larger SNR (compared to, say, a typical compact binary coalescence observed by LIGO) should help the approximation made in Eq.~\ref{eq:likeexpand} remain valid. To further establish the applicability of the Fisher matrix in this work, its results are here compared to those obtained by directly exploring the posterior density with a stochastic sampling algorithm.
 
Due to the computational cost of sampling the full posterior density, the comparison is carried out for a single EMRI around a Kerr BH with source parameters ${\log_{10}(\mu/M_\odot)=1,\,\log_{10}(M/M_\odot)=6,\,a/M=0.5,\,e_{0}=0.1,\,\cos\lambda=1/\sqrt{2}}$. These parameters were chosen to be at the approximate centre of the ranges explored in Sec.~\ref{results}, while the other parameters were the same as used in Fig.~\ref{fig.examplewave}. The event time was set to be exactly one year prior to plunge, and the luminosity distance was chosen to give an SNR of 30 for the event ($D=7.7\,\mathrm{Gpc}$ in this case).

To mitigate the computational cost associated with sampling over high-dimensional spaces, parameter estimation was performed only on the \emph{intrinsic} source parameters. The seven-dimensional intrinsic parameter space was explored using the PolyChord implementation \cite{2015MNRAS.450L..61H,2015MNRAS.453.4384H} of the nested sampling algorithm \cite{JohnSkilling}, which explores nested contours of increasing probability with a number of ``live'' points. Parameter priors were chosen to be flat over a localised volume of support, and 700 live points were used to explore this space. Highly localised prior support was required in order to facilitate sampling convergence, due to the general size and complexity of the EMRI parameter space; the prior ranges were centred on the true parameter values, and their widths were set to be five times that of the Fisher matrix estimate for the $1\sigma$ covariance contour.

Posterior slices from a PolyChord run with $5\times10^6$ likelihood evaluations are compared against the corresponding Fisher matrix predictions in Fig.~\ref{MCMCres_centralBH} (showing the properties of the central BH), Fig.~\ref{MCMCres_orbit} (showing the properties of the CO orbit), and Fig.~\ref{MCMCres_SmallBH} (showing the 1D marginalised posterior on the CO mass). As seen from these figures, the Fisher matrix method provides excellent estimates for the parameter estimation errors on all of the intrinsic source parameters. Furthermore, the PolyChord posteriors only converged after running for $\approx72\,\textrm{hours}$ on 64 cores, while the Fisher matrix was computed in $\approx 0.5\,\textrm{hours}$ on a single core.

\section{Results}\label{results}
Having established the applicability of the Fisher matrix for estimating the EMRI parameter uncertainties for the system described above, the Fisher matrix is now used to estimate the bounds it will be possible to place of the metric deformation parameters described in Sec.~\ref{GY}. To obtain these estimates the Fisher matrix is evaluated at the GR solution; i.e.\ the point where the bump parameters are zero, $\gamma_{m,n}=0$. The diagonal entries of the inverse Fisher matrix return estimates for the uncertainties on all the parameters, and the uncertainties on the each bump parameters are interpreted as an estimate of the bound that can be placed on that particular deformation. This procedure is designed to mimic the scenario where the EMRI observations are consistent with GR, and hence no deformation can be detected, and the goal is to place the most constraining limits possible. Of course, it is possible that the observations will in fact not be consistent with GR, and the goal in that case would be to measure, rather than simply constrain, the bump parameters; this possibility is not considered here.

Fisher matrices are calculated for all parameters in Eq.~\ref{eq:sys_par} (both intrinsic and extrinsic), plus one $\mathcal{B}_{N}$ bump parameter at a time. The different $\mathcal{B}_{N}$ orders are considered separately. There is no reason why bumps of different order cannot exist simultaneously (e.g.\ the dCS metric in Eq.~\ref{eq:dCSmetric} has a leading order deformation at $\mathcal{B}_{4}$ but also includes $\mathcal{B}_{5}$ and $\mathcal{B}_{6}$ terms). However, for any particular non-Kerr BH in the family described in Sec.~\ref{sec:metrics} the largest, leading order deformation will generally be the most tightly constrained by the observations.

\begin{figure}[h!]
\begin{center}
    \includegraphics[width=0.98\textwidth]{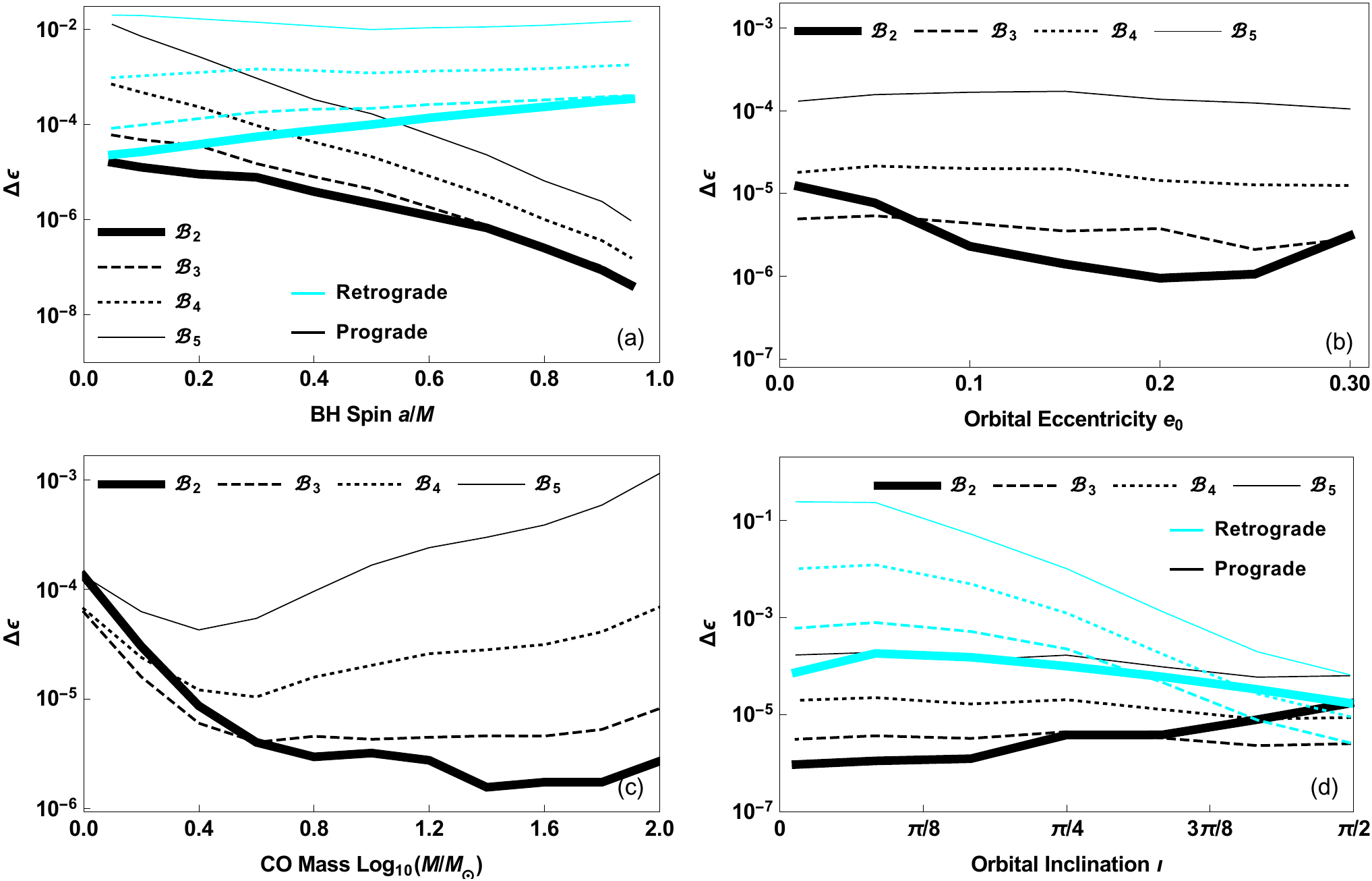}
  \caption{These plots show the Fisher matrix estimates for the bounds which may be placed on the deformation parameter $\epsilon$ for various $\mathcal{B}_{N}$ bumpy black holes given an EMRI waveform with a signal to noise ration of $\rho\!=\!30$. Panel (a) shows how the bounds vary with the spin of the central black hole for both prograde and retrograde orbits. Panel (b) shows how the bounds vary with the orbital eccentricity one year before plunge. Panel (c) shows how the bounds vary with the CO mass. Finally, panel (d) shows how the bounds vary with the orbital inclination for both prograde and retrograde orbits. Generally better bounds may be placed on lower order deformations; i.e. it is easier to constrain $\mathcal{B}_{2}$ deformations than $\mathcal{B}_{5}$. This is expected because the $\mathcal{B}_{5}$ is suppressed by higher powers of $(M/r)$ and therefore more closely mimics the Kerr metric (see Fig.\ref{fig.examplewave}). The observed trends for the four different plots are discussed in the main text. Unless otherwise indicated by the axis label the system parameters were set to default values of the EMRI system described in Sec.~\ref{sec:verify};
the central BH has mass $M=10^{6}\Msun$ and spin $S=0.7$, the CO has mass $\mu=10\Msun$ and is on an orbit with inclination $\lambda=\pi/4$, semi-major axis $a=7M$ and eccentricity $e_{0}=0.3$ at time $t=0$, and the remaining system parameters were set to ${\theta_{K}\!=\!\pi/8,\, \phi_{K}\!=\!0, \,\theta_{S}\!=\!\pi/4,\, \phi_{s}\!=\!\pi /2,\, \gamma_{0}\!=\!0, \,\Phi_{0}\!=\!0,\,\alpha_{0}\!=\!0}$.
\label{fig:mainresults}}
\end{center}
\end{figure}

The bounds on the $\mathcal{B}_{N}$ metric perturbations that are possible to place with an EMRI with SNR 30 (observed for a total of $1\,\textrm{year}$ ending at plunge) as a function of the various source parameters are shown in Fig.~\ref{fig:mainresults}. When interpreting the results of Fig.~\ref{fig:mainresults} it is helpful to keep in mind that the $\mathcal{B}_{N}$ metric perturbation is suppressed by a factor of $(M/r)^{N}$. Generally tighter constraints can be placed on the lower order deformations as these deformations are suppressed by a lower power of $(M/r)<1$.

Panel (a) shows how the bounds depend of the central BH spin. High BH spins allow for tighter constraints to be placed if the CO is on a prograde orbit because the innermost stable orbit is closer to the central BH where the metric perturbations are largest. Similar considerations explain the observed trends for retrograde orbits. 

Panel (b) shows that the constraints depend only weakly on the orbital eccentricity of the CO. High eccentricities cause the CO to approach closer to the central BH where the metric perturbations are larger, however high eccentricities also cause the CO to spend a greater proportion of its proper time are larger distances where the metric perturbations are smaller. It appears that these two effects largely serve to cancel each other out. For small eccentricities the bound on the $\mathcal{B}_{2}$ deformation degrades, this is likely due to the lack of signal power at the pericentre precession frequency leading to degeneracies with the other system parameters.

Panel (c) shows how the constraints depend on the CO mass. For heavy COs ($\mu\gtrsim50\,M_{\odot}$) the bounds degrade with increasing CO mass; this is likely due to the fact that the inspiral proceeds faster (i.e. the orbital semi-major axis changes faster) and hence a smaller fraction of the SNR is accumulated when the CO is close to the central BH. Somewhat surprisingly the bounds also degrade for light CO masses ($\mu\lesssim3\,M_{\odot}$); this is likely because in the limit of zero mass the CO remains on a single geodesic for the entire observation period and the lack of orbital evolution leads to degeneracies between the bump parameters and the other system parameters. The CO mass where the turning point occurs depends on the $\mathcal{B}_{N}$ order; for high order deformations it occurs at lower masses because the orbit must spend a significant time close to the central BH where the highly suppressed metric perturbation is significant.

Panel (d) shows how the constraints depend on the EMRI orbital inclination. The tightest constraints are obtained for prograde equatorial orbits, because it is in that case that the innermost stable orbit is closest to the central BH.

EMRI observations allow the leading order dimensionless bump parameters described in Sec.~\ref{sec:metrics} to be constrained to be less than $10^{-2}\textrm{--}10^{-7}$ depending on the system parameters. This is many order of magnitude better than can be achieved using current observations; e.g.\ observations of accretion disk in most cases cannot constrain the leading order bump parameters to be less than unity \cite{PhysRevD.92.024039}.

\section{Concluding remarks}\label{sec:discussion}
GW observations offer new possibilities for testing some of the key predictions of GR. EMRI observations with LISA are particularly well suited for addressing the question \emph{is the metric around an astrophysical supermassive black hole well described by the Kerr solution?} 
In this paper an augmented version of the widely used ``analytic kludge'' (originally proposed by \cite{BarackCutler2008}, with further improvements in \cite{2017arXiv170504259C,CG2015}) model for EMRIs around Kerr BHs has been extended to a large family of continuously parameterised deformations of the Kerr metric, known as bumpy BHs, proposed by \cite{VYS}. These bumpy BHs retain the same spacetime symmetries as the Kerr metric, namely stationarity, axisymmetry, reflection symmetry across the equatorial plane, and a second rank Killing tensor. The ``kludge'' models used here capture all of the important qualitative features of the EMRI system (including, radiation reaction, relativistic precession, orbital eccentricity etc.); however, the models are known to not remain quantitatively accurate over the multi year observation timescales for EMRIs. In future the calculations performed here may be usefully extended to more realistic EMRI models as they become available.

It has been shown that EMRIs are able to place much tighter constraints on the size of these bumps than is possible using current observations; the constraints on the dimensionless $\gamma_{m,n}$ bump parameters can improve by as much as 7 orders of magnitude. In general the best constraints will come from high SNR EMRI systems where the CO spends a large fraction of time in the very strong gravitational field; for example, a CO on a prograde orbit around a highly central BH. The size of the constraints that are possible to place using EMRI observations depend on the parameters of the EMRI and this dependence has been explored using the Fisher matrix. The validity of the Fisher matrix has been checked via a direct comparison with the full posterior probability distribution calculated using a nested sampling algorithm.

\ack{The authors would like to thank Sonke Hee for his expert advice regarding PolyChord. Part of this work was performed on the Darwin High Performance Computing Cluster at the University of Cambridge. This work has received funding from the European Union's Horizon 2020 research and innovation programme under the Marie Sk\l odowska-Curie grant agreement No.\ 690904, from STFC Consolidator Grant No.\ ST/L000636/1, and the STFC, and DiRAC's Cosmos Shared Memory system through BIS Grant No.\ ST/J005673/1 and STFC Grant Nos.\ ST/H008586/1, ST/K00333X/1.}

\section*{References}
\bibliographystyle{iopart-num}
\bibliography{bibliography}

\end{document}